\documentclass[pra,twocolumn,showpacs,groupedaddress,superscriptaddress]{revtex4}

\usepackage{amssymb}
\usepackage{graphicx}
\usepackage{bm}
\usepackage{color}
\usepackage{psfrag}
\usepackage{epstopdf}

\begin{document}

\title{A comparative study of dynamical simulation methods for the dissociation of
molecular Bose-Einstein condensates}
\author{S. L. W. Midgley}
\affiliation{The University of Queensland, School of Mathematics and Physics, ARC Centre of Excellence for Quantum-Atom Optics, Qld 4072, Australia}
\author{S. W\"{u}ster}
\altaffiliation{Current address: Max Planck Institute for the Physics of Complex Systems,
N$\ddot{\textnormal{o}}$thnitzer Strasse 38, 01187 Dresden, Germany}
\affiliation{The University of Queensland, School of Mathematics and Physics, Qld 4072, Australia}
\author{M. K. Olsen}
\author{M. J. Davis}
\author{K. V. Kheruntsyan}
\affiliation{The University of Queensland, School of Mathematics and Physics, ARC Centre of Excellence for Quantum-Atom Optics, Qld 4072, Australia}
\date{\today}

\begin{abstract}
We describe a pairing mean-field theory related to the
Hartree-Fock-Bogoliubov approach, and apply it to the dynamics of dissociation
of a molecular Bose-Einstein condensate (BEC) into correlated
bosonic atom pairs. We also perform the same simulation using two stochastic
phase-space techniques for quantum dynamics --- the positive $P$-representation
method and the truncated Wigner method. By comparing the results of
our calculations we are able to assess the relative strength 
of these theoretical techniques in describing molecular
dissociation in one spatial
dimension. An important aspect of our analysis is the inclusion
of atom-atom interactions which can be problematic for the positive-$P$ method. 
We find that the truncated Wigner method mostly agrees with the positive-$P$ simulations,
but can be simulated for significantly longer times.  The pairing mean-field theory results diverge from
the quantum dynamical methods after relatively short times.

\end{abstract}

\pacs{03.75.Nt, 03.65.Ud, 03.75.Gg, 03.75.Kk}
\maketitle

\section{Introduction}

\label{sec:intro}

%dissociation - analogue - QAOs
The dissociation of a molecular Bose-Einstein condensate (BEC) \cite{weiman,greiner,durr04,mukaiyama} into correlated atom pairs is a process
analogous to parametric down-conversion in optics. Down-conversion involving
photons has been pivotal in the advancement of quantum optics by allowing
for the generation of strongly entangled states. In the same way, molecular
dissociation has emerged as an avenue to generate strongly entangled
ensembles of atoms in the field of quantum-atom optics. This matter-wave
analog is of additional interest, however, as it gives rise to the possibility of performing tests of
quantum mechanics with mesoscopic or macroscopic numbers of massive
particles rather than with massless photons. For example, the atom pairs formed during dissociation
have Einstein-Podolsky-Rosen (EPR) type correlations in position and
momentum, and one can envisage a demonstration of the EPR paradox with
ensembles of correlated ultra-cold atoms \cite{karen, karen2, Opatrny}.
Also, molecular BECs can be formed by either two bosonic or two fermionic atoms; the latter offers the possibility of a new paradigm in fermionic quantum atom optics.

%correlations - exps. measurable
Experimental progress in the field of ultra-cold quantum gases has reached
the stage where investigation of atom-atom correlations is now possible \cite
{perrinwest, greiner}. For example, in 2005 Greiner \textit{et al.} \cite
{greiner} measured atom-atom correlations resulting from the dissociation of
$^{40}$K$_{2}$ molecules into fermionic atoms. Such advances have been
achieved through the development of techniques for the measurement of noise
in absorption images \cite{greiner,folling,altman,bach} and atom detection
using microchannel plate detectors \cite{schellekens}.

In this paper we consider correlations between bosonic atoms produced, for example, in the dissociation of $^{87}$Rb$_{2}$ dimers.
Whilst molecular dissociation of $^{87}$Rb$_{2}$ has been experimentally
realised \cite{durr04}, atom-atom correlations have not yet been measured in
these experiments due to the short molecular lifetimes. Experimental advances, however, may soon result in the production of
BECs of ro-vibrationally stable ground-state molecules \cite{lang}, in which case the present
analysis will become experimentally relevant. This paper serves to
further the understanding of atom-atom correlations in the molecular
dissociation process. Previous analytic and numeric work in this
area has been restricted to the short time limit, where the effects
of $s$-wave scattering interactions are negligible
\cite{karen1,karen06, karen05, karen, karentwin, karen2, poulsenmolmer, jack, zhao, moore2, tikhon}. However, if a
full quantitative description of atom-atom correlations is to be
obtained, the effects of spatial inhomogeneity and $s$-wave
scattering interactions on correlation strength must be addressed
\cite{karen1, karen2}. To this end, we provide numerical results
beyond the short time limit for the case of a spatially
inhomogeneous molecular condensate with atom-atom interactions
included in the model.

The other contribution made in this paper is a comparison of the performance
of three simulation methods describing the dynamics of BECs beyond Gross-Pitaevskii theory. Since the experimental realisation of Bose-Einstein condensation in 1995 \cite%
{anderson}, the dynamics of weakly interacting BECs have often been
successfully described by applying a mean-field theoretic approach leading
to the Gross-Pitaevskii equation (GPE) \cite{pethick}. However, as the GPE
neglects quantum fluctuations, its ability to describe the full BEC dynamics
is limited to cases where the effects of quantum fluctuations are negligible.

Incorporating the effects of quantum fluctuations when modelling quantum
many-body systems is necessary to describe, for example, the correlation
dynamics which play a significant role in more recent experiments, such as
molecular dissociation. As a result of this, much effort has been directed
at developing theoretical methods that go beyond mean-field theory in their
description of the dynamics of ultra-cold quantum gases \cite%
{holland,corney, murray1}. Several techniques have been used in the
analytical and numerical investigation of the dissociation of a molecular
BEC and the atom-atom pair correlations resulting from this process. For
instance, dissociation can be treated analytically using the undepleted,
classical molecular field approximation for the case of uniform condensates
\cite{karen1}; a more recent development is the analytic treatment of nonuniform condensates using a perturbation theory in time \cite%
{magnus}. As the name suggests, the undepleted molecular field
approximation assumes that the number of molecules remains constant
throughout the dissociation process. Hence, it is only valid for
short dissociation times when depletion is negligible, corresponding
to a conversion of {\color{red}}$\lesssim 10\%$ of the molecules
into atoms \cite{karen1,
matt}. Although useful in some circumstances, the obvious limitations of the analytic treatment beyond this regime necessitates an
alternative approach.

In this paper we compare three simulation techniques using molecular dissociation as an example:
two stochastic phase-space methods known as the
positive-$P$ \cite{drummond1,drummond2} and truncated Wigner 
\cite{steel98, sin1} methods, and a pairing mean-field theory known as the
Hartree-Fock-Bogoliubov (HFB) method \cite{holland2, morgan2005,
hutchinson98, griffin96}. 
%These methods go beyond the analytic treatments. 
The positive-$P$ representation
method provides an exact quantum treatment of the dissociation problem for
inhomogeneous systems, with $s$-wave scattering interactions and molecular
depletion incorporated. Extensive work has been conducted using the positive
$P$-representation method \cite%
{steel98,corney,karensupchem,poulsenmolmer,karentwin,karen,karen1,karen2,deuardrumm}%
, with Savage \textit{et al.} \cite{karen2, karen1} in particular, analysing
both position and momentum pair-correlations in molecular dissociation.

Unfortunately, the positive-$P$ approach is also limited to relatively short simulation times. For example, when
one neglects the atom-atom interactions completely, the positive-$P$
simulations are successful only for durations corresponding to about
$50\%$ conversion \cite{karen1}. For typical experimental systems, divergent trajectories
and large sampling errors arise during the evolution so that the
problem becomes intractable beyond this time scale \cite{deuar,
gilc96}. The problem becomes worse when one includes atom-atom
$s$-wave scattering interactions; in this case the dissociation
durations that can be simulated using the positive-$P$ method are
limited to only $\sim 5\%$, and at best $10\%$, conversion
\cite{karen1}. This prevents one from using the positive
$P$-representation method to determine the effects of $s$-wave scattering on the atom-atom %
correlation strength over time. Due to this limitation, there is a
subsequent lack of knowledge regarding the effects of $s$-wave scattering on
correlation dynamics for realistic condensates. This motivates further
numerical investigation of atomic correlations in molecular dissociation
using approximate methods, and to this end we consider the truncated Wigner
and HFB methods to elucidate the relative performance of the methods in the
context of molecular dissociation.

This paper is structured as follows. Section II describes the system we have
studied. Sections III and IV provide an outline of the three simulation
methods used in this work, including the relevant evolution equations and
approximations. Furthermore, it presents justification for the
use of the three methods, discusses their inherent limitations and motivates
the need for a comprehensive comparison of their relative performance. In
Section V, we detail our work based on simulations of the coupled
atom-molecule system, describing molecular dissociation in one dimension
(1D). Finally, Section VI provides an overview of work extending the
truncated Wigner simulations beyond short time scales.

\section{System and Hamiltonian}

\label{sec:hamiltonian}

We consider a molecular BEC which is dissociated into pair-correlated atoms
by way of a {magnetic Feshbach resonance. }The quantum field theory effective Hamiltonian describing this coupled atom-molecule system can be written as
\cite{karen05,karentwin},

\begin{eqnarray}
\hat{H} &=&\int d\mathbf{x}\Bigg\lbrace\sum_{i=a,m}\hat{\Psi}_{i}^{\dagger }(%
\mathbf{x})\hat{H}_{0,i}(\mathbf{x})\hat{\Psi}_{i}(\mathbf{x})  \nonumber \\
&+&\sum_{i,j=a,m}\frac{\hbar U_{ij}}{2}\hat{\Psi}_{i}^{\dagger }(\mathbf{x})%
\hat{\Psi}_{j}^{\dagger }(\mathbf{x})\hat{\Psi}_{j}(\mathbf{x})\hat{\Psi}%
_{i}(\mathbf{x})  \nonumber \\
&+&\frac{\hbar \chi }{2}\Big(\hat{\Psi}_{m}^{\dagger }(\mathbf{x})\hat{\Psi}%
_{a}^{2}(\mathbf{x})+\mathnormal{H.c.}\Big)\Bigg\rbrace  \label{hammy}
\end{eqnarray}%
where $\hat{\Psi}_{a,m}(\mathbf{x},t)$ are the atomic/molecular field
operators that annihilate an atom or molecule at position $\mathbf{x}$. The
field operators satisfy the commutation relation $[\hat{\Psi}_{i}(\mathbf{x}
,t),\hat{\Psi}_{j}^{\dagger }(\mathbf{x}^{\prime },t)]=\delta _{ij}\delta (
\mathbf{x}-\mathbf{x}^{\prime })$. The atomic/molecular free-particle
Hamiltonians, $\hat{H}_{0,a}(\mathbf{x})$ and $\hat{H}_{0,m}(\mathbf{x})$,
are given by,

\begin{eqnarray}
\hat{H}_{0,a}(\mathbf{x})&=&-\frac{\hbar ^{2}}{2m_{a}}\nabla _{\mathbf{x}%
}^{2}+\hbar V_{a}(\mathbf{x}),\\
\hat{H}_{0,m}(\mathbf{x})&=&-\frac{\hbar ^{2}}{2m_{m}}\nabla _{\mathbf{x}%
}^{2}+\hbar V_{m}(\mathbf{x})+2\hbar |\Delta |,  \label{det}
\end{eqnarray}%
where $m_{a}$ is the atomic mass and $m_{m}=2m_{a}$ is the molecular mass.
The atomic/molecular trapping potentials are given by $V_{a}(\mathbf{x})$
and $V_{m}(\mathbf{x})=2V_{a}(\mathbf{x})$. The detuning $\Delta $ in Eq.~(%
\ref{det}), or dissociation energy $2\hbar |\Delta |$, corresponds to an
overall energy mismatch of $2E_{a}-E_{m}$ between the free atom states $%
2E_{a}$ at the dissociation threshold and the bound molecular state $E_{m}$.
Hence, the process of dissociation begins with an initially stable, $%
E_{m}<2E_{a}$, molecular BEC and a magnetic field sweep onto the atomic side
of the Feshbach resonance, $E_{m}>2E_{a}$ (i.e., negative detuning $\Delta $%
), resulting in the formation of atom pairs. For a molecule at rest, the excess
dissociation energy is converted into the kinetic energy of atom pairs,
which for the most part will possess equal and opposite momenta $\pm \mathbf{
k}_{0}$, where $k_{0}=|\mathbf{k}_{0}|=\sqrt{2m_{a}|\Delta |/\hbar }$.

Returning to Eq.~(\ref{hammy}), $U_{ij}$ represents the two-body $s$-wave
interaction strengths for atom-atom, atom-molecule and molecule-molecule
scattering events. For example, $U_{aa}=4\pi \hbar a_{s}/m_{a}$ where $a_{s}$
is the atomic scattering length ($a_{s}=5.4$ nm for $^{87}$\textnormal{Rb}).
The term $\chi $ is the atom-molecule coupling and is responsible for
coherent conversion of molecules into atom pairs, where the mechanism for
conversion is via a Feshbach resonance \cite{timm, duine, kohler, drum04}. However, in
appropriately chosen rotating frames the equations can easily be recast for
conversion via optical Raman transitions \cite{karensupchem,karentwin}. In
our numerical work the atom-molecule coupling remains switched on for the
total evolution time. Also, we assume that the trapping potentials are switched off when the coupling $\chi $ is
switched on at $t=0$ and the evolution occurs in free space.

The Hamiltonian (\ref{hammy}) conserves the total number of atomic particles
\begin{equation}
N=2\langle\hat{N}_{m}(t)\rangle+\langle\hat{N}_{a}(t)\rangle=\mathrm{const},
\end{equation}%
with $\hat{N}_{i}(t)=\int d\mathbf{x}
\hat{\Psi}_{i}^{\dagger }(\mathbf{x},t)
\hat{\Psi}_{i}(\mathbf{x},t)\hspace{0.2cm}(i=a,m)$ and $N_{i}=\langle \hat{N}
_{i}\rangle $. 
We begin our simulations with the molecular BEC in a coherent state and  the
atomic field in the vacuum state, and so $ N=2\langle\hat{N}_{m}(0)\rangle$.
%, 
%where $\hat{N}_{m}(0)$ corresponds to the total initial number
% of molecules.

\section{Stochastic methods for BEC dynamics}

\label{stoch}

After being developed in the field of quantum optics, phase-space
methods have been successfully applied to matter-wave physics and have
been used in many studies of the quantum dynamics of complex many-body systems such
as BECs \cite{norrie2,deuar,poulsenmolmer,karentwin,karen1,karen06,lobo,gardiner,
steel98, drumd, drumd1, corney2, murray2, murray3, murray4}. Phase-space
representation methods rely on a mapping between the quantum operator
equations of motion and the Fokker-Planck equation (FPE) which in turn can be interpreted as a set of stochastic differential equations
(SDEs). Two distributions commonly used for this purpose can be traced back to the
Glauber-Sudarshan $P$-distribution and the Wigner distribution \cite{gardiner, walls}. Along with the HFB method, phase-space
techniques are central to this paper and hence will be discussed briefly in
order to develop a context for the numerical results presented in Sec.~V and
VI.

\subsection{The Positive $P$-representation Method}

\label{sec:posp}

The positive $P$-representation method \cite{drummond1,drummond2, deuar,
drumd, deuardrumm} enables one to perform first-principles calculations of
the quantum dynamics of multi-mode quantum many-body systems, including
BECs. It relies on exploiting the positive $P$-representation of the density
matrix, for which there exists a mapping between the master equation and a
set of c-number SDEs that can be solved numerically. The
stochastic trajectory averages calculated using the positive $P$%
-representation method correspond to the normally-ordered expectation values
of quantum mechanical operators \cite{karen1}. If stochastic sampling errors
remain small during the time evolution, any observable can, in principle,
be calculated using the positive-$P$ method.

The positive-$P$ approach requires one to double the
phase-space by defining two independent complex stochastic fields $\Psi _{i}(
\mathbf{x},t)$ and $\Phi _{i}(\mathbf{x},t)$ ($i=a,m$)
corresponding to the operators $\Psi _{i}(\mathbf{x},t)$ and $\Psi
_{i}^{\dagger }(\mathbf{x},t)$, respectively \cite{karen1}, with $\Psi 
_{i}^{\ast }(\mathbf{x},t)\neq {\Phi }_{i}(\mathbf{x},t)$ except in the mean. Using the Hamiltonian in Eq.~(\ref{hammy}), the stochastic differential
equations describing the quantum dynamical evolution are, in the appropriate
rotating frame \cite{karentwin,karen06},

\begin{eqnarray}
\frac{\partial \Psi_{a}}{\partial t} &=& \frac{i\hbar}{2m_{a}}%
\nabla^{2}\Psi_{a}-i\Big(\Delta +\sum_{i}U_{ai}\Phi_{i}\Psi_{i}\Big)%
\Psi_{a}-i\chi \Psi_{m}\Phi_{a}  \nonumber \\
&+&\sqrt{-i\chi \Psi_{m}}\zeta_{1}+\sqrt{-iU_{ma}\Psi_{a}\Psi_{m}/2}%
(\zeta_{2}+i\zeta_{3})  \nonumber \\
&+&\sqrt{-iU_{aa}\Psi^{2}_{1}}\zeta_{4},  \nonumber
\\
\frac{\partial \Phi_{a}}{\partial t} &=& -\frac{i\hbar}{2m_{a}}%
\nabla^{2}\Phi_{a}+i\Big(\Delta +\sum_{i}U_{ai}\Phi_{i}\Psi_{i}\Big)%
\Phi_{a}+i\chi \Phi_{m}\Psi_{a}  \nonumber \\
&+&\sqrt{i\chi \Phi_{m}}\zeta_{5}+\sqrt{iU_{ma}\Phi_{a}\Phi_{m}/2}%
(\zeta_{6}+i\zeta_{7})  \nonumber \\
&+&\sqrt{iU_{aa}\Phi^{2}_{1}}\zeta_{8},  \nonumber
\\
\frac{\partial \Psi_{m}}{\partial t} &=& \frac{i\hbar}{2m_{m}}%
\nabla^{2}\Psi_{m}-i \sum_{i}U_{mi}\Phi_{i}\Psi_{i}\Psi_{m}-i\frac{\chi}{2}%
\Psi^{2}_{a}  \nonumber \\
&+&\sqrt{-iU_{ma}\Psi_{a}\Psi_{m}/2}(\zeta_{2}-i\zeta_{3})+\sqrt{%
-iU_{mm}\Psi^{2}_{m}}\zeta_{9},  \nonumber
\\
\frac{\partial \Phi _{m}}{\partial t} &=&-\frac{i\hbar }{2m_{m}}\nabla
^{2}\Phi _{m}+i\sum_{i}U_{mi}\Phi _{i}\Psi _{i}\Phi _{m}+i\frac{\chi }{2}\Phi
_{a}^{2}  \nonumber \\
&+&\sqrt{iU_{ma}\Phi _{a}\Phi _{m}/2}(\zeta _{6}-i\zeta _{7})  
+\sqrt{iU_{mm}\Phi _{m}^{2}}\zeta _{10}.  \label{pospeqn}
\end{eqnarray}%
Here the  $\zeta _{j}(\mathbf{x},t)$ $(j=1,2,\,\ldots ,10)$
are real, independent, Gaussian noises with $\langle \zeta _{j}(
\mathbf{x},t)\rangle=0$ and correlations in time and space given by 
\nobreak$\langle \zeta _{j}(\mathbf{x},t)\zeta
_{k}(\mathbf{x}^{\prime },t^{\prime })\rangle =\delta _{jk}\delta
(\mathbf{x}-\mathbf{x}^{\prime })\delta (t-t^{\prime })$.

\subsection{The Truncated Wigner Method}
\label{sec:wigner}
The truncated Wigner method is another useful phase-space technique for
describing the quantum evolution of a Bose-Einstein condensate \cite%
{steel98, sin1}. Unlike the positive $P$-representation method it is an
approximate method as it involves neglecting (or truncating) third-order
derivative terms in the evolution equation for the Wigner function. This is
necessary in order to obtain an equation in the form of a Fokker-Planck
equation which can then be mapped onto a stochastic differential equation.
The third-order terms can, in principle, be represented via
stochastic difference equations, however, these are more unstable than the
positive-$P$ equations \cite{epl}. The advantage of the truncated Wigner method lies in the inclusion of
initial quantum noise, allowing the model to incorporate quantum corrections
to the classical field equations of motion and treat a different set of
problems to a Gross-Pitaevskii equation or other classical field
approaches. Although it has been shown that the truncated Wigner approach
can give erroneous results, particularly for two-time correlation functions
\cite{oc}, it can be accurate for a wide range of problems provided the particle
density exceeds the mode density \cite{sin2,norrie2}.

It can be shown that using the truncated Wigner approximation (TWA) and the
Hamiltonian in Eq.~(\ref{hammy}), the stochastic differential equations
governing the dissociation are, in the appropriate rotating frame,
\begin{eqnarray}
\frac{\partial \Psi _{a}}{\partial t}&=&\frac{i\hbar }{2m_{a}}\nabla ^{2}\Psi
_{a}-i\Big(\Delta +\sum_{i}U_{ai}|\Psi _{i}|^{2}\Big)\Psi _{a}-i\chi \Psi
_{m}\Psi _{a}^{\ast },
\nonumber\\
\frac{\partial \Psi _{m}}{\partial t}&=&\frac{i\hbar }{2m_{m}}\nabla ^{2}\Psi
_{m}-i\sum_{i}U_{mi}|\Psi _{i}|^{2}\Psi _{m}-i\frac{\chi }{2}\Psi _{a}^{2}.
\end{eqnarray}
Whilst these equations are deterministic, quantum fluctuations
are included by way of a noise contribution in the initial state for
the molecular and atomic fields. The addition of this initial vacuum
noise ensures that the initial state of $\Psi _{m}$ and $\Psi _{a}$
represent the Wigner function of an initial coherent state BEC and
an initial vacuum state, respectively. The respective
stochastic averages with the Wigner distribution function correspond
to symmetrically-ordered operator products, so that the calculation
of observables represented by normally-ordered operator products
needs appropriate symmetrisation. %As isthe case with our positive-$P$ simulations, we let $U_{am}=0$ and analyse the role of atom-atom interactions, $U_{aa}\neq 0$.

\section{Pairing Mean-Field Theory for BEC dynamics}

\label{sec:hfb}

Pairing mean-field theory -- as a simplified version of the
Hartree-Fock-Bogoliubov (HFB) approach \cite{bachey, zin, javan} -- has been applied to the problem of molecular dissociation in Refs. \cite{jack,matt}, although these works only considered spatially uniform
systems. Our present HFB study extends the analysis to nonuniform
condensates and represents the third method we use in describing the
dynamics of the molecular dissociation. This approach involves an
approximation to the full quantum evolution retaining only the lowest order
atomic fluctuations. More precisely, one writes the atomic field operator $%
\hat{\Psi}_{a}(\mathbf{x})$ in terms of the atomic mean-field $\phi _{a}(
\mathbf{x})=\langle \hat{\Psi}_{a}(\mathbf{x})\rangle $ and the lowest order
atomic fluctuations $\hat{\chi}_{a}(\mathbf{x})$, such that, $\hat{\Psi}_{a}(
\mathbf{x})=\phi _{a}(\mathbf{x})+\hat{\chi}_{a}(\mathbf{x})$. The atomic
fluctuations can be approximately represented by their lowest order correlation
functions, the normal and anomalous densities, $G_{N}(\mathbf{x},\mathbf{x}
^{\prime })=\langle \hat{\chi}_{a}^{\dagger }(\mathbf{x}^{\prime })\hat{\chi}
_{a}(\mathbf{x})\rangle $ and $G_{A}(\mathbf{x},\mathbf{x}^{\prime
})=\langle \hat{\chi}_{a}(\mathbf{x}^{\prime })\hat{\chi}_{a}(\mathbf{x}
)\rangle $, respectively \cite{holland2,wuester05}.

In our implementation the molecular field is treated as a mean-field, with $\phi_{m}(\mathbf{x})=\langle\hat{\Psi}_{m}(\mathbf{x}
)\rangle$. As suggested in Refs. \cite{holland,holland2} molecular
fluctuations can be included in the model. However, they are neglected in
our work as they are negligible on the time scales under consideration. This
is one of the main differences between the HFB and truncated Wigner
approaches, as the latter includes molecular fluctuations. By including the
fluctuation operator, $\hat{\chi}_{a}(\mathbf{x})$, the atomic field is
treated to higher order than the molecular field in our HFB formalism. This
is necessary as the atomic fluctuations play an intrinsic dynamical role in
the molecular dissociation process and also allow one to consider atomic
pair correlations. Finally, it is assumed that the initial molecular state
is a coherent state and any expectation values of greater than two atomic
fluctuation operators are factorised using Wick's theorem \cite{blaizot86},
thereby assuming that the quantum state of the system is Gaussian.

With these approximations and our Hamiltonian for the system, given in Eq.~(\ref{hammy}), we can derive a set of coupled PDEs for the atomic mean-field $
\phi _{a}(\mathbf{x})$, molecular mean-field $\phi _{m}(\mathbf{x})$ and the 
first-order correlation functions $G_{A}(\mathbf{x},\mathbf{x}
^{\prime })$ and $G_{N}(\mathbf{x},\mathbf{x}^{\prime })$. Solving these
coupled evolution equations one can then model the dynamics of dissociation of a molecular BEC.

Within the HFB formalism the evolution equations describing the molecular
dissociation process are given by
\begin{eqnarray}
\frac{\partial \phi _{a}(\mathbf{{x})}}{\partial t} &=&\frac{i\hbar }{2m_{a}}
\nabla _{\mathbf{x}}^{2}\phi _{a}(\mathbf{x}) \nonumber\\
&&-iU_{aa}\left[ |\phi _{a}(\mathbf{x})|^{2}+2G_{N}(\mathbf{x},\mathbf{x})%
\right] \phi _{a}(x) \nonumber\\
&&-iU_{aa}G_{A}(\mathbf{x},\mathbf{x})\phi _{a}^{\ast }(\mathbf{x})-i\chi
\phi _{m}(\mathbf{x})\phi _{a}^{\ast }(\mathbf{x}),\nonumber\\
\\
\frac{\partial \phi_{m}(\mathbf{{x})}}{\partial t}&=&\frac{i\hbar}{4m_{a}}%
\nabla^{2}_{\mathbf{x}}\phi_{m}(\mathbf{x})-2i| \Delta|\phi_{m}(\mathbf{x})
\nonumber \\
&&-iU_{mm}|\phi_{m}(\mathbf{x})|^{2}\phi_{m}(\mathbf{x}) -i\frac{\chi}{2}%
[\phi_{a}^{2}(\mathbf{x})\nonumber\\
&&+G_{A}(\mathbf{x},\mathbf{x})],  
\label{hfbeqn2}
\end{eqnarray}

\begin{eqnarray}
\frac{\partial G_{A}(\mathbf{x},\mathbf{x}^{\prime})}{\partial t}&=&-\frac{i%
}{\hbar}\langle [\hat{\chi}(x^{\prime})\hat{\chi}(x),\hat{H}]\rangle
\nonumber \\
&=&\frac{i\hbar}{2m_{a}}\nabla^{2}_{\mathbf{x}}G_{A}(\mathbf{x},\mathbf{x}%
^{\prime}) +\frac{i\hbar}{2m_{a}}\nabla^{2}_{\mathbf{{x}^{\prime}}}G_{A}(%
\mathbf{x},\mathbf{x}^{\prime})  \nonumber \\
&-&2iU_{aa}\Big[|\phi_{a}(\mathbf{x})|^{2}+|\phi_{a}(\mathbf{x}%
^{\prime})|^{2}+G_{N}(\mathbf{x},\mathbf{x})  \nonumber \\
&+&G_{N}(\mathbf{x}^{\prime},\mathbf{x}^{\prime})\Big]G_{A}(\mathbf{x},%
\mathbf{x}^{\prime})  \nonumber \\
&-&iU_{aa}\Big[\phi_{a}(\mathbf{x})^{2}G_{N}^{*}(\mathbf{x},\mathbf{x}%
^{\prime})+\phi_{a}(\mathbf{x}^{\prime})^{2}G_{N}(\mathbf{x},\mathbf{x}%
^{\prime})  \nonumber \\
&+&G_{A}(\mathbf{x},\mathbf{x})G_{N}^{*}(\mathbf{x},\mathbf{x}^{\prime})
+G_{A}(\mathbf{x}^{\prime},\mathbf{x}^{\prime})G_{N}(\mathbf{x},\mathbf{x}%
^{\prime})\Big]  \nonumber \\
&-&iU_{aa}\Big[\phi_{a}(\mathbf{x})^{2}+G_{A}(\mathbf{x},\mathbf{x})\Big]%
\delta(\mathbf{x}-\mathbf{x}^{\prime})  \nonumber \\
&-&i\chi\Big[\phi_{m}(\mathbf{x})[ G_{N}^{*}(\mathbf{x},\mathbf{x}%
^{\prime})+\delta(\mathbf{x}-\mathbf{x}^{\prime})]  \nonumber \\
&+&\phi_{m}(\mathbf{x}^{\prime})G_{N}(\mathbf{x},\mathbf{x}^{\prime})\Big],
 \label{hfbeqn3}
\end{eqnarray}
\begin{eqnarray}
\frac{\partial G_{N}(\mathbf{x},\mathbf{x}^{\prime })}{\partial t} &=&-\frac{%
i}{\hbar }\langle \lbrack \hat{\chi}^{\dagger }(x^{\prime })\hat{\chi}(x),%
\hat{H}]\rangle  \nonumber  \label{hfbeqn4} \\
&=&\frac{i\hbar }{2m_{a}}\nabla _{\mathbf{x}}^{2}G_{N}(\mathbf{x},\mathbf{x}%
^{\prime })-\frac{i\hbar }{2m_{a}}\nabla _{\mathbf{{x}^{\prime }}}^{2}G_{N}(%
\mathbf{x},\mathbf{x}^{\prime })  \nonumber \\
&-&2iU_{aa}\Big[|\phi _{a}(\mathbf{x})|^{2}-|\phi _{a}(\mathbf{x}^{\prime
})|^{2}+G_{N}(\mathbf{x},\mathbf{x})  \nonumber \\
&-&G_{N}(\mathbf{x}^{\prime },\mathbf{x}^{\prime })\Big]G_{N}(\mathbf{x},%
\mathbf{x}^{\prime })  \nonumber \\
&-&iU_{aa}\Big[\phi _{a}(\mathbf{x})^{2}G_{A}^{\ast }(\mathbf{x},\mathbf{x}%
^{\prime })+\phi _{a}^{\ast }(\mathbf{x}^{\prime })^{2}G_{A}(\mathbf{x},%
\mathbf{x}^{\prime })  \nonumber \\
&+&G_{A}(\mathbf{x},\mathbf{x})G_{A}^{\ast }(\mathbf{x},\mathbf{x}^{\prime
})-G_{A}^{\ast }(\mathbf{x}^{\prime },\mathbf{x}^{\prime })G_{A}(\mathbf{x},%
\mathbf{x}^{\prime })\Big]  \nonumber \\
&-&i\chi \Big[\phi _{m}(\mathbf{x})G_{A}^{\ast }(\mathbf{x},\mathbf{x}%
^{\prime })-\phi _{m}^{\ast }(\mathbf{x}^{\prime })G_{A}(\mathbf{x},\mathbf{x%
}^{\prime })\Big],  \nonumber \\
&&
\end{eqnarray}%
where $G_{N}(\mathbf{x},\mathbf{x})$ is the density of the noncondensed
atoms. This follows from the expression for the total density of atoms, $
\langle \hat{\Psi}_{a}\hat{\Psi}_{a}\rangle =|\phi _{a}|^{2}+\langle \hat{
\chi}_{a}^{\dagger }\hat{\chi}_{a}\rangle $, where $|\phi _{a}|^{2}$ is the
density of the condensate atoms and $\langle \hat{\chi}_{a}^{\dagger }\hat{
\chi}_{a}\rangle $ is the density of the noncondensed atoms.

In our simulations we neglect $\phi _{a}(\mathbf{{x})}$ as the atomic field
does not develop since all the terms on the right-hand side of Eq.~(\ref{hfbeqn2}) are multiplied by the field or its conjugate. The physics here is similar to that of a non-degenerate optical parametric oscillator with phase diffusion \cite{reid}. More
particularly, only the sum of the phases of each correlated atom pair is
known, fixed to the phase of the molecular BEC, whilst the relative phase is
unknown and takes an arbitrary value. It follows that the individual phases
of these correlated modes are also arbitrary and consequently, no common
phase to the atomic field exists across the entire range of momenta.
%Finally, as in our stochastic simulations, we let $U_{am}=0$ and investigate the role of $U_{aa}\neq 0$.

The potential role of the HFB and truncated Wigner methods, despite being
approximate techniques, is to model realistic inhomogeneous condensates in
which the effects of $s$-wave scattering interactions on atom-atom pair
correlations can be quantified and compared with experimental data.
Moreover, both methods present the possibility of describing physics in
regimes for which the positive-$P$ representation method is computationally
intractable. Prior to this work, however, there
has been no attempt to undertake a comprehensive comparison of the
performance of all three of these methods when applied to the same problem, although there has been comparisons of the positive-$P$ and truncated Wigner methods when investigating BEC collisions \cite{deuued}. This
motivates the comparative study of these approximate methods and the
positive-$P$ approach. 

There are further potential advantages in developing the HFB method for
application to such problems. The positive-$P$ representation and truncated Wigner methods require the averaging of many trajectories
(corresponding to quantum mechanical ensemble averaging) and therefore
requires multiple runs. In contrast, since the HFB method is
not a stochastic technique it only necessitates a single run but
at the expense of the dimensions of the problem being doubled.
Also, in many ways it is a more intuitive method, with the derivation
highlighted here being an extension of the well-known Gross-Pitaevskii
approach.

\section{Comparison of positive-$P$, HFB and Truncated Wigner Results}

\label{sec:compare}

\subsection{Parameter values}

\label{sec:yh}

We now present 1D simulations for the
dissociation of a $^{87}\textnormal{Rb}_{2}$ molecular BEC with
$m_{a}=1.44\times 10^{-25}$ kg and $m_{m}=2m_{a}$. For computational simplicity we consider an effective one-dimensional (1D) system by assuming strong harmonic confinement in the transverse direction. All parameters are
chosen to be close to a typical experimental system, with the
exception of a relatively small value for the detuning $|\Delta |$ so that the computational grid need not be too large. In practice, the
detuning should be such that the total dissociation energy $2\hbar
|\Delta |$ is much larger than the thermal energy due to the finite
temperature of the system; here, we assume a zero-temperature
condensate. At the same time the detuning $|\Delta |$ should be
smaller than the frequency of the transverse harmonic trap
potential, so that transverse excitations are suppressed and the
dynamics of dissociation remains in 1D. 

We set $U_{mm} = 0$ and $U_{am} = 0$ in our simulations; it is the role of atom-atom s-wave scattering that is of particular interest in this work, and to this end we perform simulations with both $U_{aa} = 0$ and $U_{aa} \ne 0$. Setting $U_{mm}$ to be zero is unlikely to be entirely physical, however for a more realistic value we find that the positive-$P$ simulations become intractable after a very short time, making our goal of a comparison impossible.  Additionally, we find that there is no significant difference between the TWA simulation for $U_{mm} = 0$ and $U_{mm} \neq 0$, and so this has no practical implications for our study.

Similar considerations apply to the atom-molecule interactions which
are set to $U_{am} = 0$. At the mean-field level, the atom-molecule
interactions in the equations of motion for the atomic field would
initially appear as an effective spatially varying detuning that
depends on the molecular BEC density profile; these interactions can
be neglected if the total dissociation energy $2\hbar |\Delta|$ is
much larger than the respective interaction energy per atom. For our
choice of $\Delta$ and the molecular BEC peak density (see below)
this would in turn require an atom-molecule scattering length of
$\lesssim 0.1$ nm. For more realistic values of the atom-molecule
scattering length (assumed to be in the few nanometers range) the
approximation would require absolute detunings in the kHz range or
higher and it would improve with increasing $|\Delta|$.

%Neglecting the atom-molecule interaction can be justified if the respective interaction energy per atom is much smaller than the total dissociation energy $2\hbar|\Delta|$. Since the atomic density is small initially and the overlap with the molecular cloud becomes smaller as the dissociation progresses, the approximation is indeed reasonable and it improves with increasing $|\Delta|$.

The initial molecular BEC density is taken to be gaussian,
\begin{equation}
n_{m}(x,t=0)=n_{0}e^{-x^{2}/\sigma^{2}},
  \label{ggf}
\end{equation}%
corresponding to a trapping frequency of 0.15 Hz in the x-direction with a harmonic oscillator length of 50 $\mu$m, where $n_{0}=1.83\times 10^{7}$ m$^{-1}$ is the molecular BEC peak 1D (linear) density. The size of the
one-dimensional quantisation box was chosen to be $L=6.5\times
10^{-4}$ m and the lattice grid was composed of $512$ points. The atom-atom
interaction strength is given by $U_{1D}=U_{aa}/2A=2\omega
_{\bot }a_{s}$ \cite{karen2005}, where $A=\pi l_{\bot
}^{2}$ is the confinement area in the transverse direction, with $l_{\bot }=
\sqrt{\hbar /m\omega _{\bot }}$ being the transverse ground-state harmonic
oscillator length and $\omega _{\bot }$ is the transverse oscillation
frequency. The atom-molecule coupling in 1D is given by $\chi _{1D}=\chi 
\sqrt{2\pi l_{\perp }^{2}}=6.7\times 10^{-3}$ m$^{1/2}$ s$^{-1}$ \cite{matt},
and is switched on for the total evolution time. The detuning $\Delta
=-258 $ s$^{-1} $ and hence, $k_{0}=\sqrt{2m_{a}|\Delta |/\hbar }=8.41\times
10^{5} $ m$^{-1}$ is the resonant momentum at $\pm k_{0}$.

With the reduction of the coupling constants to
their 1D counterparts, the equations of motion in previous sections
are unchanged, except that all propagating fields (and the
respective noise sources in the positive-$P$ equations) are now
understood as 1D fields, while the
operator $\nabla _{\mathbf{x}}^{2}$ is replaced by $\partial ^{2}/
\partial x^{2}$.

In all our simulations, we assume that the atomic field is initially in a
vacuum state and that the molecular condensate is initially in a coherent
state, with density profile given by Eq.~(\ref{ggf}). The trapping potential
is turned off when the dissociation coupling $\chi_{1D}$ is
switched on, with a Feshbach sweep into the dissociation regime $\Delta<0$.
Distinct from the implementation of the stochastic methods, the HFB
simulations assume that the molecular condensate remains in a coherent state
during the dynamical evolution. Also, the initial atomic fluctuation fields
are assumed to be $G_{A}(\mathbf{x},\mathbf{x}^{\prime})=G_{N}(\mathbf{x},
\mathbf{x}^{\prime})=0 $ and the molecular fluctuations are omitted.

The numerical codes for solving the evolution equations for the methods
under consideration, were implemented using the {\small XMDS} simulation
package \cite{xmds}. All stochastic simulations were performed for the case
of 10,000 trajectories. In the following sections, we have verified the
accuracy of the results presented by ensuring, for instance, invariance of
results for different lattice size and time step. Furthermore, we were able
to perform benchmarking with the analytic result within the undepleted,
molecular field approximation up until $t= 0.06 $s $(\sim 10 \%$ molecular
conversion), and more importantly, with the exact positive-$P$ results.

\subsection{Initial comparisons}

\label{sec:uj}

We first perform simulations neglecting atom-atom interactions with $%
U_{1D}=0$ for an initial number of molecules $N_{m}(0) = 1.62 \times 10^{3}$. We observe the formation of peaks in the atomic density at
momenta $\pm k_{0}$ as the dissociation energy (excess potential energy) is
converted into the kinetic energy of the correlated atom pairs, with equal
but opposite momentum $\pm k_{0}$. We verify that the value of $k_{0}$
agrees with the predicted value, given in Sec.~\ref{sec:yh}.

\begin{figure}[tbhp]
\includegraphics[width=0.45\textwidth]{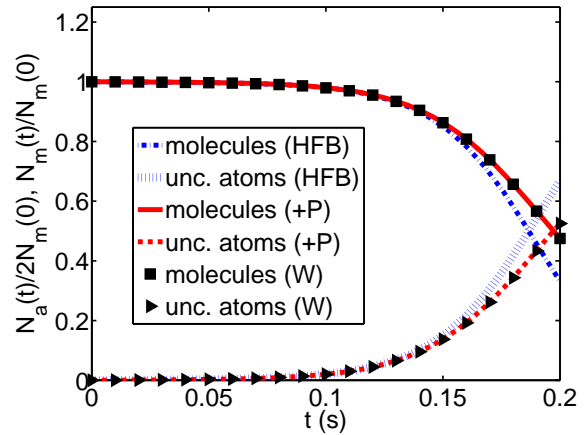}
\caption{(Color online) Comparison of the fractional particle numbers for $%
t_{{\protect\tiny \mathnormal{final}}}$ = 0.20 s, for the case of a
non-uniform condensate with $U_{1D}=0$, for positive-$P$ (red
solid and dashed lines), truncated Wigner (black $\blacksquare$ and $\blacktriangleright$) and HFB (blue
dashed lines) methods. The fractional atom numbers (initially lower curves) and
the fractional molecule numbers are shown. In this figure and throughout this paper, the positive-$P$ and truncated
Wigner results are for the case of 10,000 trajectories. The error bars are
shown and are essentially the thickness of the data lines in all figures.
In all simulations performed the initial number of molecules is $%
N_{m}(0) = 1.62 \times 10^{3}$. }
\label{fig:image1}
\end{figure}

In Fig.~\ref{fig:image1} we provide a plot of the total fractional number of
molecules $N_{m}(t)/N_{m}(0)$ and the total fractional number of atoms $%
N_{a}(t)/2N_{m}(0)$ as a function of time $t$, normalised to the total
molecule $N_{m}(0)$ and total atom number $2N_{m}(0)$, respectively.
Although this result does not include the effects of $s$-wave scattering, it
does allow one to compare the performance of the methods. It can be seen
that all the methods agree until $t\sim 0.14$ s, which corresponds to the
conversion of $\sim 10$\% of the molecules. It is found that whilst the
truncated Wigner method does extremely well when compared to the exact
results provided by the positive-$P$ method, the HFB method diverges
substantially at longer times. 

The positive-$P$ method becomes intractable at $t \sim 0.20$ s, and hence
the ability to compare all three methods ceases beyond this point. Looking
forward, when one incorporates $s$-wave scattering the positive-$P$ method
will fail sooner \cite{karen1} and hence the truncated Wigner and HFB
methods may be able to access a regime otherwise inaccessible to numerical
simulations for realistic non-uniform condensates.

\subsection{Observation of Phase Diffusion Processes During Dissociation of
a Molecular BEC}

\label{sec:diffusion}

\begin{figure}[tbhp]
\includegraphics[width=0.45\textwidth]{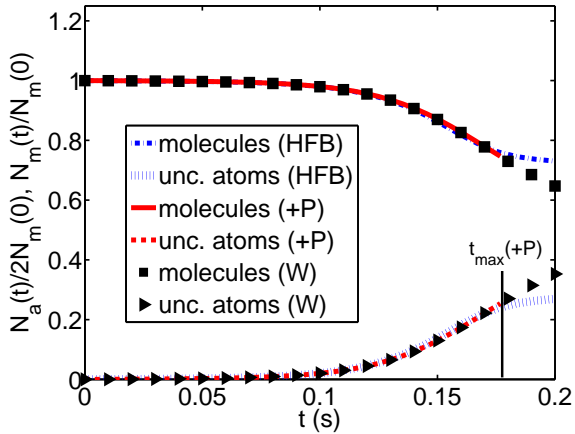}
\caption{(Color online) As in Fig.~\protect\ref{fig:image1}, except for the
case of a non-uniform condensate with $U_{1D}= g_0$. The
positive-$P$ method becomes intractable beyond $t_{{\mathnormal%
{\protect\tiny max}}}$(+P) $\sim0.18$ s, compared with $t\sim0.2$ s when $s$
-wave scattering is neglected. We also find that the number of molecules
converted into atom pairs is decreased when $s$-wave scattering interactions
are included and attribute this to phase diffusion.}
\label{fig:image2}
\end{figure}

In this section we consider non-uniform condensates with $s$-wave scattering
interactions included. In Fig.~\ref{fig:image2}, we plot the fractional
particle number throughout the evolution, for the same parameters as in Fig.~\ref{fig:image1}, but with scattering included. We choose an interaction strength of 
$
U_{1D}\equiv g_0 = 1.04\times 10^{-6} \omega_{\perp}a_{s}$, which corresponds to 
$^{87}$Rb with transverse confinement of $\omega _{\bot }/2\pi = 30$ Hz and s-wave scattering length of $a_s=5.4$ nm.
From these results it can be
seen that the positive-$P$ method fails beyond approximately $t_{\max }(+
\mathnormal{P})=0.18$ s, whilst the truncated Wigner method produces results
beyond $t_{\max }(+P)$ and still does well in comparison to the positive-$P$
results up to $t_{\max }(+P)$. As expected \cite{karen1}, we find that the
positive-$P$ method fails for even shorter times as the interaction strength
is increased. For example, with an interaction strength of 
$
U_{1D} = 32 g_0$  we find that the positive-$P$
method fails for $t_{\max }(+\mathnormal{P})\sim 0.05$ s.

Here we also observe the signature of phase diffusion during molecular
dissociation \cite{lewen96,steel98,xiong2002,li2001}. By considering Eq.~(\ref{pospeqn}) we are able to estimate the characteristic %
diffusion time for the dissociation process,
\begin{equation}
t_{d}\sim \frac{\pi }{2U_{1D}\langle \hat{\Psi}^{\dagger }(x=0,t_{d})\hat{%
\Psi}(x=0,t_{d})\rangle},
\end{equation}%
and verify that the process of phase diffusion is responsible for
suppressing molecular conversion, and hence, decreasing the number
of atom pairs formed. We also performed simulations for increased
values of the atom-atom interaction strength, $U_{1D}=2 g_0$ and $U_{1D}= 32 g_0$. The results show
that molecule conversion decreases with increasing interaction
strength and further supported the order of magnitude estimates of
the diffusion time. Unfortunately, as in Sec.~\ref{sec:uj}, we see
that the HFB method still fails to adequately describe the dynamics of the
molecular dissociation process for longer times, with the particle
numbers only providing bounds for the true values. Another feature
indicative of the limitations of the HFB method is its inability to
predict the formation of peaks in the molecular momentum
spectrum at $\pm 2k_{0}$ \cite{magnuspriv}, in addition to the main resonant momenta peaks formed at $\pm k_{0}$.
These secondary peaks arise due to atom-atom recombination processes $k_{0}+k_{0}\rightarrow 2k_{0}$ and $-k_{0}-k_{0}\rightarrow
-2k_{0}$, and are observed in the positive-$P$ and truncated Wigner results. They do not arise in
the HFB results as the method does not allow for uncondensed molecules outside
the initial condensate mode.

\subsection{Analysis of Atomic Pair-Correlation Functions}

\label{sec:correlations}

We have also investigated atomic pair-correlations resulting from the
dissociation process, for realistic non-uniform condensates including the
effects of $s$-wave scattering. The strength of atom-atom correlations can
be quantified using Glauber's second-order correlation function $g^{(2)}(%
\mathbf{k},\mathbf{k}^{\prime},t)$ \cite{glauber},

\begin{equation}
g^{(2)}(\mathbf{k},\mathbf{k}^{\prime },t)=\frac{\left\langle \hat{a}^{\dagger }%
(\mathbf{k},t)\hat{a}^{\dagger }(\mathbf{k}^{\prime},t)\hat{a}(\mathbf{k}^{\prime},t)\hat{a}(\mathbf{k},t)\right\rangle }{%
\left\langle \hat{n}(\mathbf{k},t)\right\rangle \left\langle \hat{n}(\mathbf{k}^{\prime},t)\right\rangle },  \label{eqq}
\end{equation}%
with the momentum-space density at $\mathbf{k}$ given by $n(\mathbf{k},t)=\left\langle\hat{n}(\mathbf{k},t)\right\rangle =\left\langle\hat{a}%
^{\dagger }(\mathbf{k},t)\hat{a}(\mathbf{k},t)\right\rangle$ where the momentum-space field amplitudes are represented by the lattice-discretized momentum components $\hat{a}^{\dagger}(\mathbf{k})$ and $\hat{a}(\mathbf{k})$, which correspond to the continuous Fourier transforms of the fields in the limit $ \Delta k \rightarrow 0$ \cite{karen1}. This pair-correlation function describes the ratio of the probability of
the joint detection of atom pairs with $\mathbf{k}$ and $\mathbf{{k}^{\prime
}}$ to the product of the probabilities of independent atom detection at $%
\mathbf{k}$ and $\mathbf{{k}^{\prime }.}$ From this it follows that $g^{(2)}(%
\mathbf{k},\mathbf{k}^{\prime },t)=1$ for uncorrelated atom pairs, $g^{(2)}(%
\mathbf{k},\mathbf{k}^{\prime },t)=2$ for thermally bunched atoms and $%
g^{(2)}(\mathbf{k},\mathbf{k}^{\prime },t)>2$ for strongly correlated atoms
\cite{karen1}.

\begin{figure}[tbhp]
\centering
\includegraphics[width=0.45\textwidth]{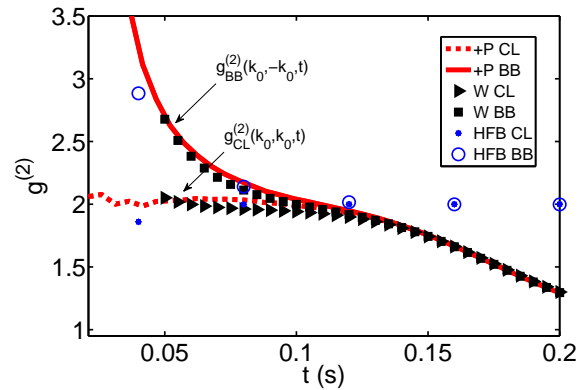}
\caption{(Color online) Plot of the atomic pair-correlation functions for
back-to-back and collinear scattering processes, denoted $%
g^{(2)}_{BB}(k_{0},-k_{0},t)$ and $g^{(2)}_{CL}(k_{0},k_{0},t)$. Results are
shown for $t_{\mathnormal{\protect\tiny final}}$ = 0.20 s, for a non-uniform
condensate with $U_{1D}=0$. The positive-$P$ results (red solid and dashed lines), the
truncated Wigner results (black $\blacksquare$ and $\blacktriangleright$) and the HFB results (blue $\circ$ and
$\ast$) are shown. The collinear (red dashed, black $\blacktriangleright$ and blue $\ast$) and
the back-to-back pair-correlations (red solid, black $\blacksquare$ and blue $%
\circ $) are shown. }
\label{fig:image3}
\end{figure}

We quantify pair-correlations arising due to momentum conservation which are
present between atoms with equal but opposite momenta, and pair-correlations
arising due to quantum statistical effects [i.e. the Hanbury-Brown and Twiss
(HBT) bunching] between atoms scattered in the same direction. The atomic
pair-correlations function for back-to-back (BB) and collinear (CL)
scattering processes, are denoted $g_{BB}^{(2)}(k_{0},-k_{0},t)$ and $%
g_{CL}^{(2)}(k_{0},k_{0},t)$, respectively. These quantities are shown in
Fig.~\ref{fig:image3} for the case of no $s$-wave scattering and in Fig.~\ref%
{fig:image4} with scattering incorporated. The collinear correlation
indicates HBT thermal bunching with $g_{CL}^{(2)}(k_{0},k_{0},t)=2$ until $%
t\sim 0.10$ s in both cases. The back-to-back correlation is super-bunched
due to strong correlations between atom pairs with equal but opposite
momenta, with $g_{BB}^{(2)}(k_{0},-k_{0},t)>2$ for short times. Beyond $%
t\sim 0.10$ s we observe both the collinear and back-to-back correlations
are approximately equal, fall below two, and approach the uncorrelated 
or coherent level of $g^{(2)}=1$, for the truncated Wigner and
positive-$P$ results. The HFB results, on the other hand, fail to predict
where $g^{(2)}$ approaches the coherent state level as stimulated processes
become important. Toward the end of the simulation, the back-to-back
correlation drops below the collinear correlation, $%
g_{BB}^{(2)}(k_{0},-k_{0},t)<g_{CL}^{(2)}(k_{0},k_{0},t)$. This effect
becomes more severe with increasing values of $U_{1D}$ and is also 
noticeable in the HFB results. %iesqz lost beyond this point

\begin{figure}[tbhp]
\centering
\includegraphics[width=0.45\textwidth]{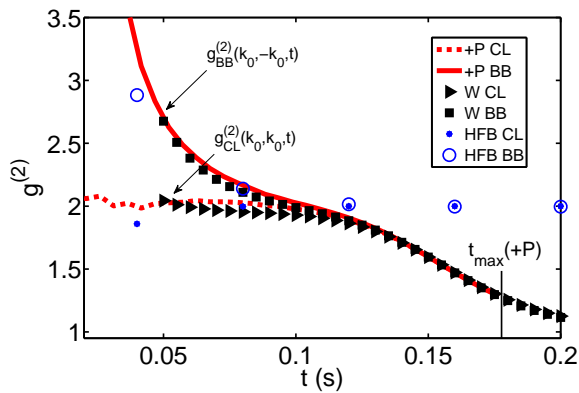}
\caption{(Color online) As in Fig.~\protect\ref{fig:image3}, except for the
case of a non-uniform condensate with $U_{1D}=g_0$. The
positive-$P$ method becomes intractable beyond $t_{{\mathnormal%
{\protect\tiny max}}}$(+P)$\sim0.18$ s, compared with $t\sim0.2$ s when $s$
-wave scattering is neglected. By comparison with Fig.~\protect\ref{fig:image3}, we see that the back-to-back and collinear pair-correlation
strength degrades at an increased rate when $s$-wave scattering is included.}
\label{fig:image4}
\end{figure}

From Figs. \ref{fig:image3} and \ref{fig:image4} it is again 
clear that the truncated Wigner method is most successful in describing
molecular dissociation, with the positive-$P$ method intractable at longer
times. It should be noted that the truncated Wigner results for this correlation function are not shown
prior to $t\sim 0.05$ s, where sampling issues arise due to the small number of atoms per mode. However, once the signal is significant the results agree with positive-$P$. As seen in Sec.~V B and C, the HFB method
fails to fully describe the dynamics for longer times as the
molecular field deviates from the assumed coherent state.

\section{Simulations beyond $t_{\max }(+P)$}

\label{sec:long times}

\begin{figure}[tbhp]
\includegraphics[width=0.43\textwidth]{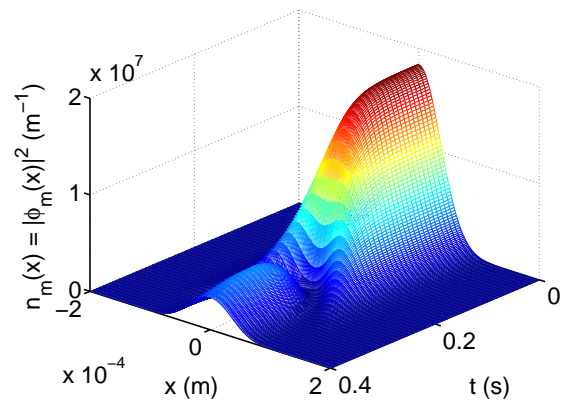}
\caption{(Color online) Molecular density in position space $n_{m}(x)$ [in
units of m$^{-1}$] as a function of time for the
HFB results. The molecular peak density is given by $n_{0} = 1.83 \times
10^{7} $m$^{-1}$.}
\label{fig:imagef}
\end{figure}

The numerical results we have presented indicate that the HFB method is
unsuitable for quantitative correlation studies of molecular dissociation
beyond the regime of the positive-$P$ simulations. The HFB method becomes
invalid as it assumes a mean-field coherent state for molecules for the
entire simulation time \cite{jack,matt}. However, once molecular depletion
reaches $\sim 80\%$, this assumption is no longer valid and the method
becomes increasingly inadequate as the regime of complete depletion is
reached. This assertion is further supported in Fig.~\ref{fig:imagef}, which
provides a surface plot of the molecular density in position space beyond $%
t_{\max }(+P)$. Here we begin to observe the development of ripple effects which coincide
with the reduction of the molecular condensate density and it is unlikely that the approximation of the molecular field as a coherent state is still valid. In this regime the effects of quantum fluctuations become
increasingly important and hence, we cannot rely on the HFB method. To
remedy this, the inclusion of molecular fluctuations $\hat{\chi}_{m}$ in the
HFB formalism could be one avenue for future work. It should be stressed
that there is value in using the HFB method, as it lies between the crude
undepleted, molecular field approximation and an exact quantum treatment.
For instance, the HFB approach is suited to high energy, sparsely occupied
modes \cite{blakie2007}. In such cases, fluctuation effects are largely
insignificant and the HFB method is valuable.

\begin{figure}[tbhp]
\includegraphics[width=0.45\textwidth]{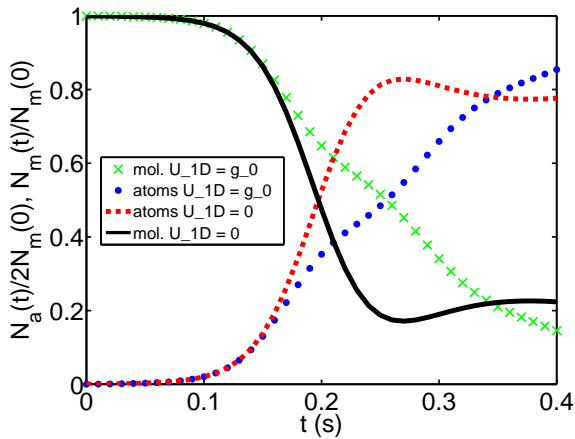}
\caption{(Color online) Fractional particle numbers for $t_{{\protect\tiny \mathnormal{final%
}}}$ = 0.40 s, for the case of a non-uniform condensate with $U_{1D}=0$ (black solid line and red dashed line) and $U_{1D}=g_0$ (green crosses and blue dotted), for
the truncated Wigner results. The fractional atom number (blue dotted and red dashed)
and the fractional molecule number (black solid and green crosses) are shown. For the $U_{1D}=0$ case for $t >$
0.20 s, we observe an increase in the molecular population from atom-atom recombination. With the
inclusion of $s$-wave scattering interactions we again see the effects of
phase diffusion.}
\label{fig:image8}
\end{figure}

\begin{figure}[tbhp]
\includegraphics[width=0.45\textwidth]{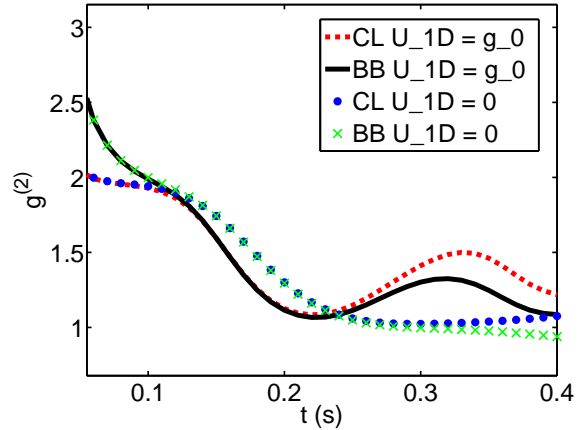}
\caption{(Color online) Plot of the atomic pair-correlation functions for back-to-back (black solid and green crosses) and
collinear (blue dotted and red dashed) scattering processes, $g^{(2)}_{BB}(k_{0},-k_{0},t)$ and $%
g^{(2)}_{CL}(k_{0},k_{0},t)$, respectively. Results are shown for $t_{
\mathnormal{\protect\tiny final}} = 0.40$ s, for a non-uniform condensate
with $U_{1D}=0$ (blue dotted and green crosses) and $U_{1D}=g_0$ (black solid and red dashed). The back-to-back correlation
drops below the collinear correlation. }
\label{fig:image7}
\end{figure}

With the HFB approach found to be invalid beyond the realm of the positive-$%
P $ simulations, we look at extending the simulations using the truncated
Wigner approach. Fig.~\ref{fig:image8} and \ref{fig:image7} repeat the analysis given in Sec.~\ref{sec:compare} but with
the extension to $t=0.40$ s. In Fig.~\ref{fig:image8} we again look at the
fractional particle numbers for the cases neglecting and including atom-atom interactions.
Beyond $t \sim 0.2$ s, for the $U_{1D}=0$ case we observe the effects of atom-atom
recombination. This is apparent due to the slight increase in the
number of molecules, and corresponding decrease in the number of atoms,
until $t \sim 0.3$ s. With the effects of $s$%
-wave scattering included, we again witness a phase diffusion process which
is responsible for decreasing the rate of molecule conversion.

In Fig.~\ref{fig:image7} we provide the truncated
Wigner results for the back-to-back and collinear pair-correlation functions
for the cases where $s$-wave scattering interactions are neglected and
included. In both cases, the back-to-back correlation is exceeded by
the collinear correlation, i.e. $%
g_{BB}^{(2)}(k_{0},-k_{0},t)<g_{CL}^{(2)}(k_{0},k_{0},t)$, with the effect becoming more dramatic with time. It is interesting to note that when atom-atom interactions are neglected the back-to-back pair correlation eventually turns into anti-correlation, ie. $g^{(2)}_{BB}(k_{0},-k_{0})<1$, for sufficiently long 
times when the molecular depletion is large. As the atomic density increases, we see atom-atom recombination which is not correlated at the two momenta considered, so that atoms are not removed equally from each of the modes under consideration. Overall, by using the truncated Wigner
method to go beyond the realm of the positive-$P$ simulations, we are able
to observe the effects of $s$-wave scattering on correlation dynamics for
realistic inhomogeneous condensates.

\section{Conclusions}

In this work we have compared three different theoretical approaches to the
problem of dissociation of molecular Bose-Einstein condensates. We have
considered the case where the atoms resulting from this dissociation process
are not trapped, but move away from the parent molecules with momenta that
are a function of the detuning. In particular, we have calculated atomic and
molecular populations and analysed the effects of atom-atom interactions
beyond the short time limit for inhomogeneous condensates. We have also
investigated quantum correlations, providing quantitative results for the
back-to-back and collinear pair-correlations, which cannot be calculated in
the standard mean-field Gross-Pitaevskii approach. This is a subject of
immediate interest as experiments which can measure these correlations can
now be performed, particularly with metastable helium \cite{perrinwest}.

In principle, the preferred theoretical method would be the stochastic
integration of equations in the positive-$P$ representation, as these give
complete access to all properties of the interacting many-body quantum
system. In practice, however, the problems inherent in the integration of
these equations, especially when $s$-wave interactions of any appreciable
strength are present, mean that the positive-$P$ equations are only
useful for short times. Another method which has been widely
applied to model BEC dynamics is the HFB approach. In some sense this is
equal to the commonly used linearisation procedures of quantum optics, and
similarly to that area, we find that we must be careful with its validity.
In fact, we have shown here that the HFB approach will sometimes become
inaccurate on shorter time scales than those which give problems in the
positive-$P$ representation approach. Although it does present computational
advantages in that the equations need only be solved once by contrast with
the phase-space representations where averages need to be taken over many
realisations, we see that it is also not useful for all parameter regimes.
This could be remedied, at least in part, by including the effects of
molecular fluctuations. However, this is a cumbersome and computationally
expensive process. 

We have found that the most useful of the methods is the truncated Wigner
representation. Although the approximations necessary to obtain stochastic
differential equations mean that the mapping from the quantum Hamiltonian is
not exact, we find that the truncated Wigner method agrees with the
first-principle positive-${P}$ results whenever such a comparison is
possible to make. It also has the advantages of not suffering from the
stability problems of the positive-$P$ representation and is valid over
longer times than the HFB approach. In conclusion therefore, we find that
while the positive-$P$ and HFB approaches are useful in some regimes, the
truncated Wigner representation is best suited to this problem.

\section*{Acknowledgments}

This work was supported by the Australian Research Council Centre of Excellence for
Quantum-Atom Optics (CE0348178), the ARC Discovery Project scheme (DP0343094), and an award
under the Merit Allocation Scheme of the National Facility of the Australian
Partnership for Advanced Computing.

\bibliographystyle{apsrev}
\bibliography{hfbbibpaper}

\end{document}